\documentclass[10pt,twocolumn,letterpaper]{article}
\usepackage{booktabs}
\usepackage{multirow}

\usepackage{diagbox}
\usepackage{iccv}
\usepackage{authblk}
\usepackage{times}
\usepackage{epsfig}
\usepackage{graphicx}
\usepackage{amsmath}
\usepackage{amssymb}
\usepackage{enumitem}
\usepackage{multirow}
\usepackage{booktabs}
\usepackage{bbm}
\usepackage[T1]{fontenc}
\usepackage[utf8]{inputenc}
\usepackage[english]{babel}
\usepackage[autostyle, english=american]{csquotes}
\usepackage{subcaption}

\MakeOuterQuote{"}
\usepackage{bbm}

\usepackage[breaklinks=true,bookmarks=false]{hyperref}

\iccvfinalcopy 

\def\iccvPaperID{****} 

\ificcvfinal\fi
\ificcvfinal\fi
\begin{document}

\title{Solution for Temporal Sound Localisation Task of ECCV Second Perception Test Challenge 2024}


\author{

Haowei Gu\textsuperscript{1},
Weihao Zhu\textsuperscript{1},
Yang Yang\textsuperscript{1, $\thanks{Corresponding Author}$} 
}

\affil{

 $^1$Nanjing University of Science and Technology
}
\maketitle

\begin{abstract}

 This report proposes an improved method for the Temporal Sound Localisation~(TSL) task, which localizes and classifies the sound events occurring in the video according to a predefined set of sound classes. The champion solution from last year's first competition has explored the TSL by fusing audio and video modalities with the same weight. Considering the TSL task aims to localize sound events, we conduct relevant experiments that demonstrated the superiority of sound features~(Section~\ref{exp}). Based on our findings, to enhance audio modality features, we employ various models to extract audio features, such as InterVideo, CaVMAE, and VideoMAE models. Our approach ranks first in the final test with a score of 0.4925.

\end{abstract}

\section{Introduction}

Deep learning techniques attract widespread attention in multiple research fields~\cite{0074ZGGZ22, YangHGXX23, YangWZX019, YangZZXJY23, YangWZL018, YangYBZZGXY23, YangFZLJ21, YangZZX019}. Integrated into sound localization, which focuses on locating and following a specific sound source across consecutive frames, they play a critical role in computer vision for tasks such as scene understanding and action recognition. It entails classifying sound events from both video and audio modalities.

Localizing and classifying the sound events occurring in the video according to a predefined set of classes,
known as temporal sound localization~(TSL), remain a challenging problem in video understanding. In this competition, we use the champion solution  Temporal Sound Localisation Task of ICCV 1st Perception Test Challenge 2023 as the baseline.
Significant progress has been made in developing deep models for TSL. For example, TallFormer~\cite{cheng2022tallformer} extracts a video representation with a Transformer-based encoder.

Currently, these methods have greatly improved many problems in sound localization, but there is still much to explore. 
For example, the champion solution from last year's 1st competition pays the same attention to the audio and video features. Since TSL task aims to localize sound events, we propose that the audio play a more important role in it. Our subsequent experiments confirme this hypothesis. 
To address it, we find that to better locate the sound events, it is important to assign more weight to audio features. Thus we employ various models to extract audio features, such as InterVideo~\cite{wang2022internvideo}, CAV-MAE~\cite{gong2022contrastive}, and VideoMAE~\cite{tong2022videomae} models to enhance audio features.

\begin{figure*}[ht]
    \centering
    \includegraphics[width=1.0\textwidth,height=0.5\textwidth]{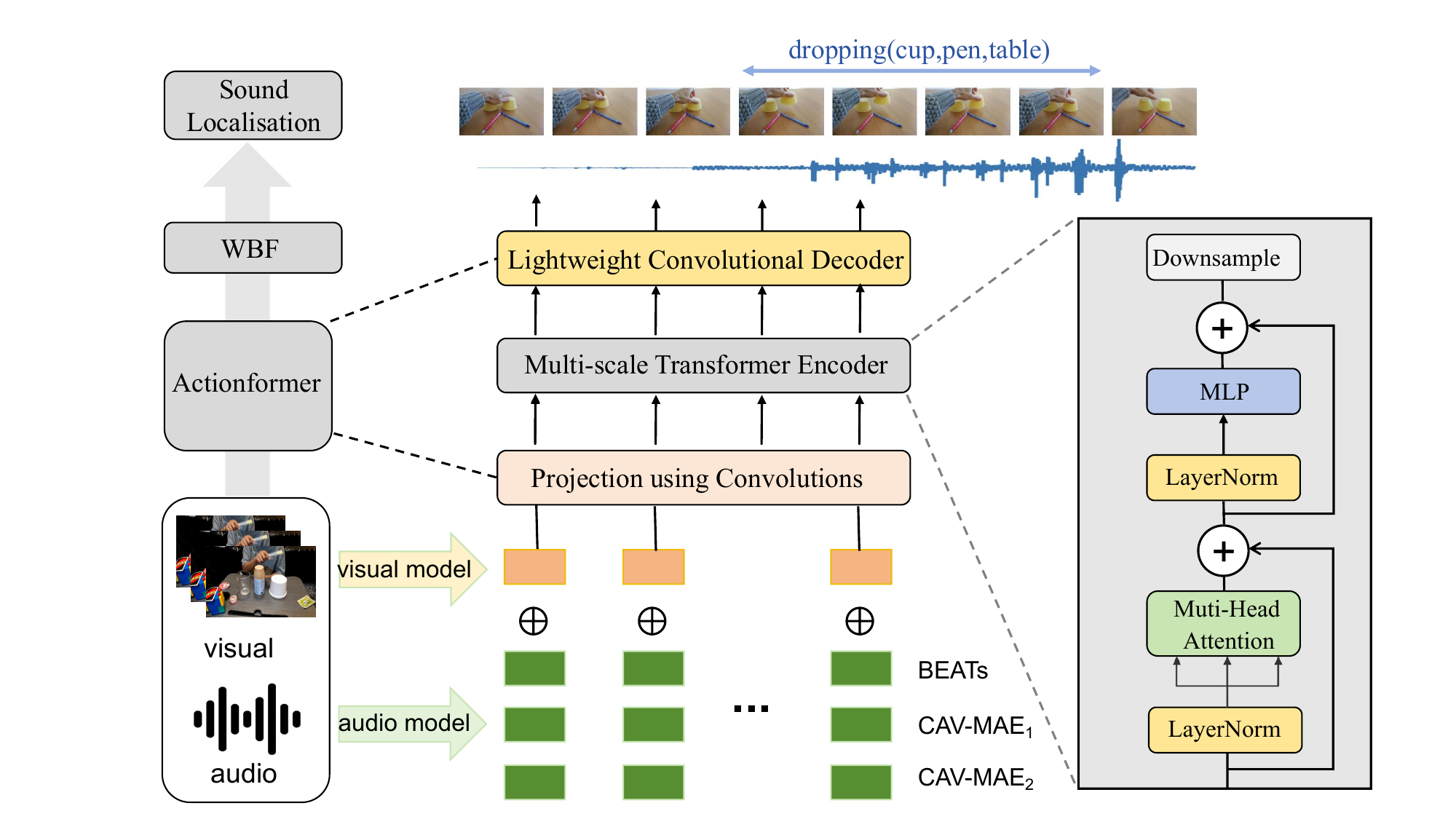}
    \caption{This network utilizes a multimodal fusion approach for action detection. Video features are extracted using the UMT model, while audio features are generated from the BEATS model and two variants of CAV-MAE, fine-tuned on AudioSet and VGGSound, respectively. The audio outputs are concatenated to form a comprehensive audio representation. Subsequently, the fused video and audio features are processed by the Actionformer network, with the final action detection results refined through a post-processing step using Weighted Box Fusion.}
    \label{fig: problem}
\end{figure*}

\section{Method}
\subsection{Preliminary}
The sound localization task requires localizing and classify the sound events occurring in the video according to a predefined set of sound classes. Given an untrimmed audio
$A=\{a_{t}\}_{t=1}^{T}$, where T denotes the audio length and t denotes the t-th frame, the sound events are
defined as $\Psi_{g} = \psi_{n}=(t_{s}, t_{e}, c)_{n=1}^{N_{g}}$. $t_{s}$, $t_{e}$, and c indicates the start, end, and category on the n-th
sound events of the audio instance $\psi$, respectively, and $N_{g}$ is the total events number of the current
audio $A_{g}$.

\subsection{Overall Architecture}
As shown in Figure \ref{fig: problem}, we adopt Actionformer~\cite{zhang2022actionformer} as the foundational model, integrating multi-scale feature representations with local self-attention. The model utilizes a lightweight decoder to classify individual moments and estimate corresponding sound boundaries. For multimodal input, visual features are derived from the fine-tuned UMT-Large~\cite{li2023unmasked} model and the VideoMAE-large model. To enhance the audio modality and achieve a more comprehensive acoustic representation, we leverage three models: the BEATS~\cite{chen2022beats} model and two variants of the CAV-MAE model fine-tuned on AudioSet~\cite{gemmeke2017audio} and VGGSound~\cite{chen2020vggsound}, respectively. Audio features are first extracted independently from each of these models and then concatenated to provide a richer audio representation, which is subsequently fused with the video features to form a robust multimodal input.

\subsection{Multimodal Feature Extraction}
For the video features, we employ two models: the fine-tuned UMT-Large model and the VideoMAE-Large model. Both models have achieved state-of-the-art performance across a range of downstream video tasks.
 
 For the audio features, we adopt a multi-model strategy to construct a comprehensive acoustic representatio. We utilize three state-of-the-art pre-trained models: BEATS and two variants of CAV-MAE fine-tuned on AudioSet and VGGSound, respectively. All audio inputs are first transformed into mel-spectrograms, serving as the foundation for subsequent feature extraction.Each model independently processes these mel-spectrograms to extract high-level features. The dimension of each audio feature is 768. We then concatenate the three feature vectors, yielding a 2304-dimensional representation that encapsulates a wide range of acoustic characteristics.
 
To ensure coherent multimodal integration, we perform temporal alignment between video and audio features using interpolation techniques. We then apply an early fusion strategy, where video and audio features are concatenated along the channel dimension, resulting in a unified multimodal representation that enables effective cross-modal learning and temporal context modeling.

\subsection{Temporal Sound Localization}
We utilize Actionformer as the baseline model for action localization to predict sound categories and boundaries. The integrated video and audio features are encoded into a feature pyramid through a multi-scale Transformer for subsequent processing. The resulting feature pyramid is then passed through shared regression and classification heads to generate sound candidates at each time step.

\subsection{Post-Processing}
 Weighted Boxes Fusion~(WBF)~\cite{solovyev2021weighted} is commonly employed to combine predictions from object detection models. We have made modifications to the WBF method to ensure its applicability to the TSL task.
 
We extract video features using the UMT model and the VideoMAE model separately and concatenate these with audio features for model training. We generate results from models trained for 20, 25, and 30 epochs for each method. Subsequently, these results were input into the WBF module, with identical fusion weights assigned to each model's results.

\section{Experiment}
\label{exp}
\textbf{Dataset.} The dataset used in Perception Test Challenge track3 includes videos~(RGB+audio) that are high resolution and have 17 sound events. We only rely on official train data to train our model, without any additional data. Both the validation and test sets use the official data provided.


\textbf{Metric.} The mean average precision~(mAP)~\cite{everingham2010pascal} is the evaluation metric used for this challenge. It is determined by calculating the average precision across various action classes and IoU~\cite{DBLP:conf/aaai/ZhengWLLYR20} thresholds. We evaluate the IoU thresholds in increments of 0.1, ranging from 0.1 to 0.5.

\textbf{Experimental Results.} 
Table \ref{tab: compare} presents a performance comparison,  demonstrating the sound features significantly enhance the overall effectiveness of the approach. These features provide valuable insights and allow for better sound localization.




\begin{table}[htp]
    \centering
    \begin{tabular}{cc}
    \hline
    \ {Model}&{map}  \\
    \hline
    Baseline & 0.3925    \\
    Cavmae(Sound only)  & 0.4560 \\
    VideoMae(Video only)  & 0.4102   \\
    All & 0.4710 \\
    WBF  & 0.4925   \\
    \toprule
    \end{tabular}
\caption{Ablation experiment.Cavmae(Sound only) means we only use extra Sound features extracted from Cavmae model.VideoMae(Video only)  means we only use extra Video features. All means using both VideoMae video features and Cavmae sound features. }
\label{tab: compare}

\end{table}

\section{Conclusion}
In this report, we employ the champion solution  Temporal Sound Localisation Task of ECCV Second Perception Test Challenge 2024 as the baseline. We use various models to extract video and sound features and pay more attention to the sound modality. In the end, we achieve a test set mAP of 0.4925, ranking first on the leaderboard.

{\small
\bibliographystyle{ieee_fullname}
\bibliography{PT_new}
}
\end{document}